\documentclass[12pt]{article}

\usepackage{scicite}
\usepackage{graphicx}
\usepackage{times}
\PassOptionsToPackage{hyphens}{url}\usepackage{hyperref}
\usepackage{xcolor}
\usepackage{subfiles}
\usepackage{amsmath,booktabs,siunitx}
\usepackage{makecell}
\usepackage{fixltx2e}
\usepackage{hyperref}
\usepackage{vcell}
\newcommand{\norm}[1]{\left\lVert#1\right\rVert}
\newenvironment{sciabstract}{%
\begin{quote} \bf}
{\end{quote}}

\title{AI-powered multimodal modeling of personalized hemodynamics in aortic stenosis}

\author
{Caglar Ozturk,$^{1,2\dag}$ Daniel H. Pak,$^{3\dag}$ Luca Rosalia,$^{1,4}$\\ Debkalpa Goswami,$^{5,6}$ Mary E. Robakowski,$^{5,7}$ Raymond McKay,$^{8}$ \\ 
Christopher T. Nguyen,$^{5,6}$
James S. Duncan,$^{3\ast}$
Ellen T. Roche$^{1,4,9\ast}$
\\
\\
\normalsize{$^{1}$ Institute for Medical Engineering and Science, Massachusetts Institute of Technology,}\\ 
\normalsize{Cambridge, MA, USA}\\
\normalsize{$^{2}$ Bioengineering Science Research Group, School of Engineering, University of Southampton,}\\ 
\normalsize{ Southampton, UK}\\
\normalsize{$^{3}$ Department of Biomedical Engineering, Yale University, New Haven,
CT, USA}\\
\normalsize{$^{4}$ Health Sciences and Technology Program, Harvard–Massachusetts Institute of Technology,}\\
\normalsize{Cambridge, MA, USA}\\
\normalsize{$^{5}$ Cardiovascular Innovation Research Center, Heart, Vascular \& Thoracic Institute,}\\
\normalsize{ Cleveland Clinic, Cleveland, OH, USA}\\
\normalsize{$^{6}$ Department of Cardiovascular Medicine, Heart, Vascular \& Thoracic Institute,}\\
\normalsize{ Cleveland Clinic, Cleveland, OH, USA}\\
\normalsize{$^{7}$Department of Chemical and Biomedical Engineering, Cleveland State University,}\\
\normalsize{Cleveland, OH, USA}\\
\normalsize{$^{8}$ Interventional Cardiology, Hartford Hospital,
Hartford, CT, USA}\\
\normalsize{$^{9}$ Department of Mechanical Engineering, Massachusetts Institute of Technology,}\\ 
\normalsize{Cambridge, MA, USA}\\
\\
\normalsize{$^\ast$Corresponding authors. E-mails:  james.duncan@yale.edu (JSD); etr@mit.edu (ETR)}\\
\normalsize{$^\dag$These authors contributed equally to this work.}
\normalsize{ computational fluid dynamics; aortic stenosis}
}

\date{}

\begin{document}

% Double-space the manuscript.

\baselineskip24pt

% Make the title.

\maketitle 
\clearpage
% Place your abstract within the special {sciabstract} environment.
\begin{sciabstract}
Aortic stenosis (AS) is the most common valvular heart disease in developed countries. High-fidelity preclinical models can improve AS management by enabling therapeutic innovation, early diagnosis, and tailored treatment planning. However, their use is currently limited by complex workflows necessitating lengthy expert-driven manual operations. Here, we propose an AI-powered computational framework for accelerated and democratized patient-specific modeling of AS hemodynamics from computed tomography. First, we demonstrate that our automated meshing algorithms can generate task-ready geometries for both computational and benchtop simulations with higher accuracy and 100 times faster than existing approaches. Then, we show that our approach can be integrated with fluid-structure interaction and soft robotics models to accurately recapitulate a broad spectrum of clinical hemodynamic measurements of diverse AS patients. The efficiency and reliability of these algorithms make them an ideal complementary tool for personalized high-fidelity modeling of AS biomechanics, hemodynamics, and treatment planning. 

\end{sciabstract}

\section*{Introduction}

Aortic stenosis (AS) is a progressive pathological condition characterized by the narrowing of the aortic valve (AV) orifice. In AS, calcium often accumulates on the valve leaflets, affecting the opening and closing of the AV \cite{ramaraj2008degenerative,chambers2009aortic}. AS is the prevailing valvular heart condition in developed countries, affecting 2-5\% of adults over 65 years old \cite{lindroos1993prevalence,freeman2004acquired,charlson2006decision,lee2023esc}. Due to the aging population, these figures are projected to triple over the next few decades \cite{danielsen2014prevalence}. As a result of increasing the left ventricular (LV) afterload, AS not only manifests as symptoms including dyspnea, syncope, and angina, but may also initiate a cascade of heart abnormalities \cite{marquis2016medical,benfari2018concomitant,slimani2020relative}. Recent guidelines have recommended early management of AS to reduce major adverse cardiovascular and cerebrovascular events \cite{lee2023esc}. The growing prevalence of AS and the heterogeneity in the anatomy and hemodynamics associated with AS underscore the need for high-fidelity models that can enhance the management of each individual patient \cite{weinberg2009computational,miller2011calcific}.

High-fidelity patient-specific models have the potential to improve the standard-of-care for AS diagnosis and treatment, as current diagnostic methods $-$ largely based on ultrasound imaging $-$ may underestimate cardiovascular risks in patients with only mildly abnormal hemodynamics \cite{carabello2009aortic,otto2020acc,brennan2012long,jm2001repeat}. Diagnostically, computational models can complement these methods by analyzing alternative sources of patient data and providing additional diagnostic information \cite{bouvier2006diagnosis,pflederer2010aortic,saikrishnan2014accurate}. Therapeutically, physical and digital twins can help predict risks (e.g., stroke, paravalvular leak, and coronary obstruction) associated with current interventional and surgical approaches, such as transcatheter aortic valve replacement (TAVR). Therefore, these models have the potential to improve patient outcome via device optimization and personalized treatment planning \cite{dowling2019patient,bosi2020validated,rotman2018realistic,hosny2019pre}.

Computational geometric reconstruction from patient images is essential for high-fidelity patient-specific modeling of AS. Due to the complex structure of calcified AV, existing methods have been mostly limited to static single-frame reconstruction using sequential refinements from voxelgrid segmentation and parametric surfaces \cite{morganti2014simulation,luraghi2019modeling,hosny2019pre,levin20203d}. With this sequential approach, extensive manual post-processing is often required for each downstream analysis, leading to significant challenges in reproducibility as well as in the widespread adoption of AS models. This limitation is further exacerbated for dynamic multi-frame modeling, which traditionally involves subsequent feature-based registration or computational simulations with active material models \cite{veress2005measurement,baillargeon2014living,gurev2015high}. Some methods have circumvented the reconstruction steps by utilizing simplified parameterization \cite{wald2018numerical,dede2021computational}, ad-hoc input flow \cite{youssefi2017patient,kimura2017patient}, or highly tunable soft robotic systems \cite{rosalia2022soft,rosalia2023soft}. However, the full 3D geometry of the relevant heart structures is still required to recapitulate the detailed blood flow characteristics of AS, such as flow vorticity and TAVR-induced blood stasis \cite{hope2010bicuspid,ducci2013hemodynamics,vahidkhah2017valve}.

Here, we present a computational framework for personalized modeling of AS hemodynamics with detailed 3D patient geometry (Fig.~1). The proposed fully automated meshing algorithms empowered by artificial intelligence (AI) significantly enhance the speed and accuracy of AS geometry reconstruction, effectively addressing the scalability and practical limitations of conventional engineering approaches. We demonstrate the versatility of our framework through multimodal modeling, including static, dynamic, computational, and benchtop analyses. For computational modeling, our fluid-structure coupled multi-physics model enables precise simulations of the valve leaflet motion and the resulting blood flow at a high temporal resolution. Using only cardiac computed tomography (CT) angiography as the input patient data, our \textit{in silico} and \textit{in vitro} models accurately reproduce the clinical continuous-wave (CW) Doppler measurements of hemodynamics for AS patients. Our framework may help accelerate the development and usage of highly precise models for AS hemodynamics in research and clinical practice.

\begin{figure}[hbt!]
     \centerline{\includegraphics[width=0.88\textwidth,trim=0 5cm 0 1cm ,clip]{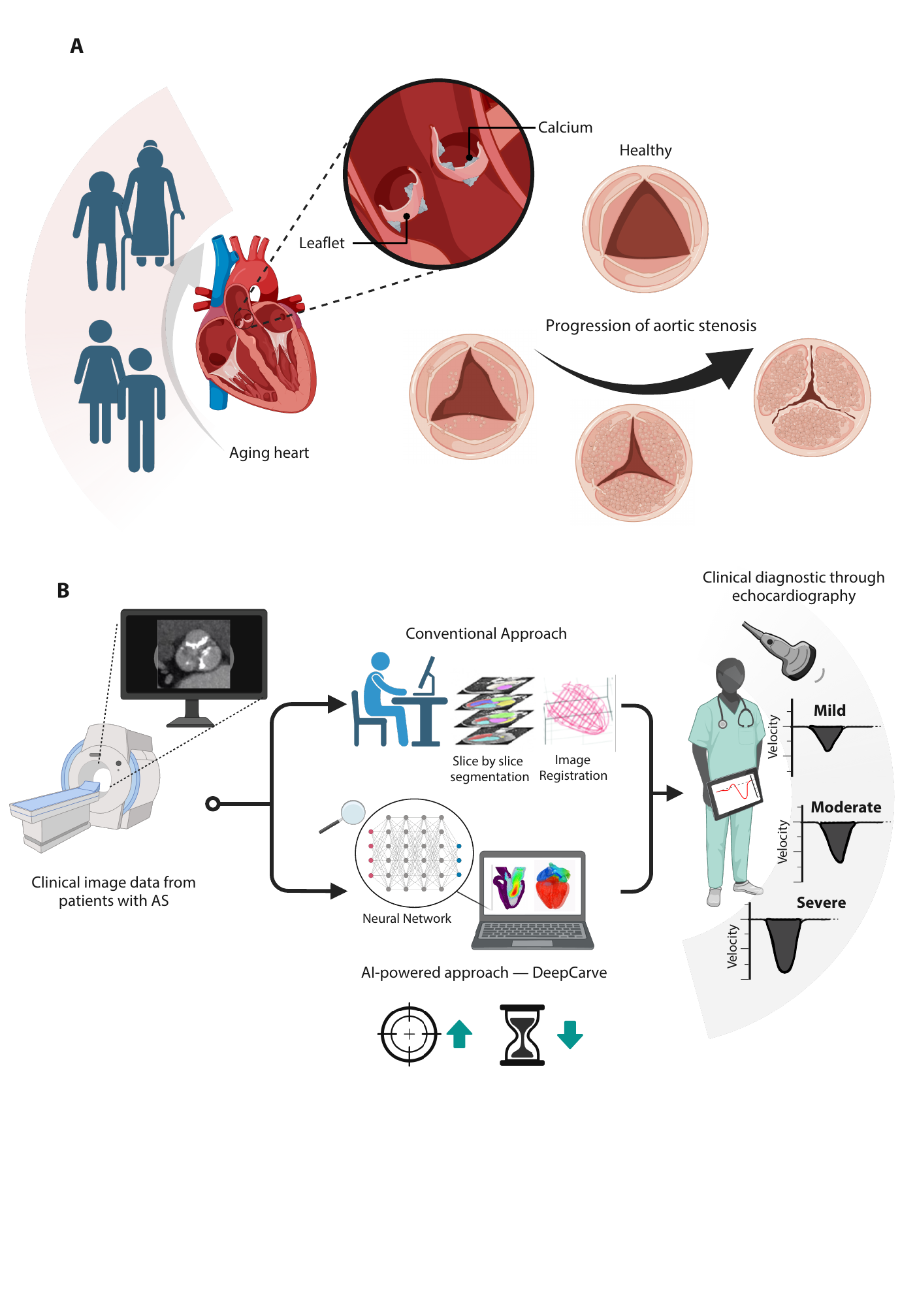}}
    {\bf Fig. 1. }{\textbf{Overview of AI-powered personalized modeling approach for AS.} \textbf{(A)} Schematic illustrating the calcification process of the AV leaflets in the aging heart and changes in the orifice area through disease progression. \textbf{(B)} The overall workflow of developing a patient-specific model from CT images to recapitulate clinical hemodynamic measurements of AS patients, where we propose the use of our DeepCarve-FSI framework to replace conventional heart meshing techniques.}
    \label{fig:overview}
\end{figure}

% Our framework consists of data-driven AI meshing algorithms and physics-driven fluid-structure coupled models.
% As a complementary solution to the recent engineering innovations
% directly addresses the current limitations in scalable modeling of precise AS geometry
% which makes sense because cardiac CT is already being used more and more for calcium scoring and coronary flow and CT is also used for essentially every TAVR pre-procedural planning. 
% comprising the left ventricle, ascending aorta, aortic valve leaflets, and calcification. 

\section*{Results}

\subsection*{AI-powered geometry reconstruction for computational AS modeling}

% Top-down approach: simulation mesh to all others
% main differences: whole surface LV and tri-layer leaflets, improved metrics from original DeepCarve
% DeepCarve and C-MAC overall description
% static and dynamic

We leveraged AI to directly generate the most complex finite element (FE) representation of AS geometry $-$ volumetric meshes. From this detailed mesh, we can automatically derive all other necessary representations, ensuring high-quality inputs with accurate multi-part relationships for various modeling modalities. Our AI-powered volumetric meshing involves two stages: heart tissue reconstruction and calcium deposit attachments. We have previously developed related algorithms for both stages, namely DeepCarve \cite{pak2023patient} and C-MAC \cite{pak2024robust}, but their use cases have been limited to static computational structural analyses. Here, we extend these algorithms to provide a holistic computational framework for multimodal modeling, and provide the first set of validations against clinical measurements of AS hemodynamics.

DeepCarve constitutes an image-processing neural network that can rapidly reconstruct simulation-ready volumetric meshes from pre-TAVR CT (Movie S1). The original model was designed for the aortic valvular complex and partial LV with minimized solid element distortions, which often led to conservative deformation and imprecise LV boundaries. This is especially problematic for the elderly AS population, who often exhibits unusual LV volume due to concurrent cardiomyopathies or remodeling processes secondary to AS. For dynamic flow analysis, precise delineation of the entire LV boundary is necessary to accurately define the boundary of the fluid domain. Here, we therefore re-defined the LV mesh to entirely consist of surface elements and introduced surface element quality metrics in the training loss. Further, we modified the valve leaflet meshes to contain three layers of hexahedral elements in the thickness direction, which increases the accuracy of the highly dynamic leaflet movements. \cite{sun2014computational,qin2024patient}.

% Evaluating the spatial accuracy of the modified DeepCarve against the ground truth and a conventional registration method.

We performed geometric evaluations of the AI-generated meshes via quantitative measurements (Table.~1) and qualitative visualizations (Fig.~2A). We focused on evaluating the spatial accuracy as well as the element quality of the reconstructed meshes. Compared to conventional registration, which consists of a deep learning segmentation algorithm followed by segmentation-based template registration (Methods; Fig. S1, Fig. S2, Movie S2), our model showed statistically significant improvements in all evaluation metrics for all structural components. The performance differences are also visually apparent, especially along the LV endocardium and the outer aortic wall. In later sections, we will demonstrate how these geometric improvements lead to increased downstream task performance in hemodynamic modeling.
\begin{figure}[hbt!]
    \centerline{\includegraphics[width=1\textwidth,trim=0 0 0 0 ,clip]{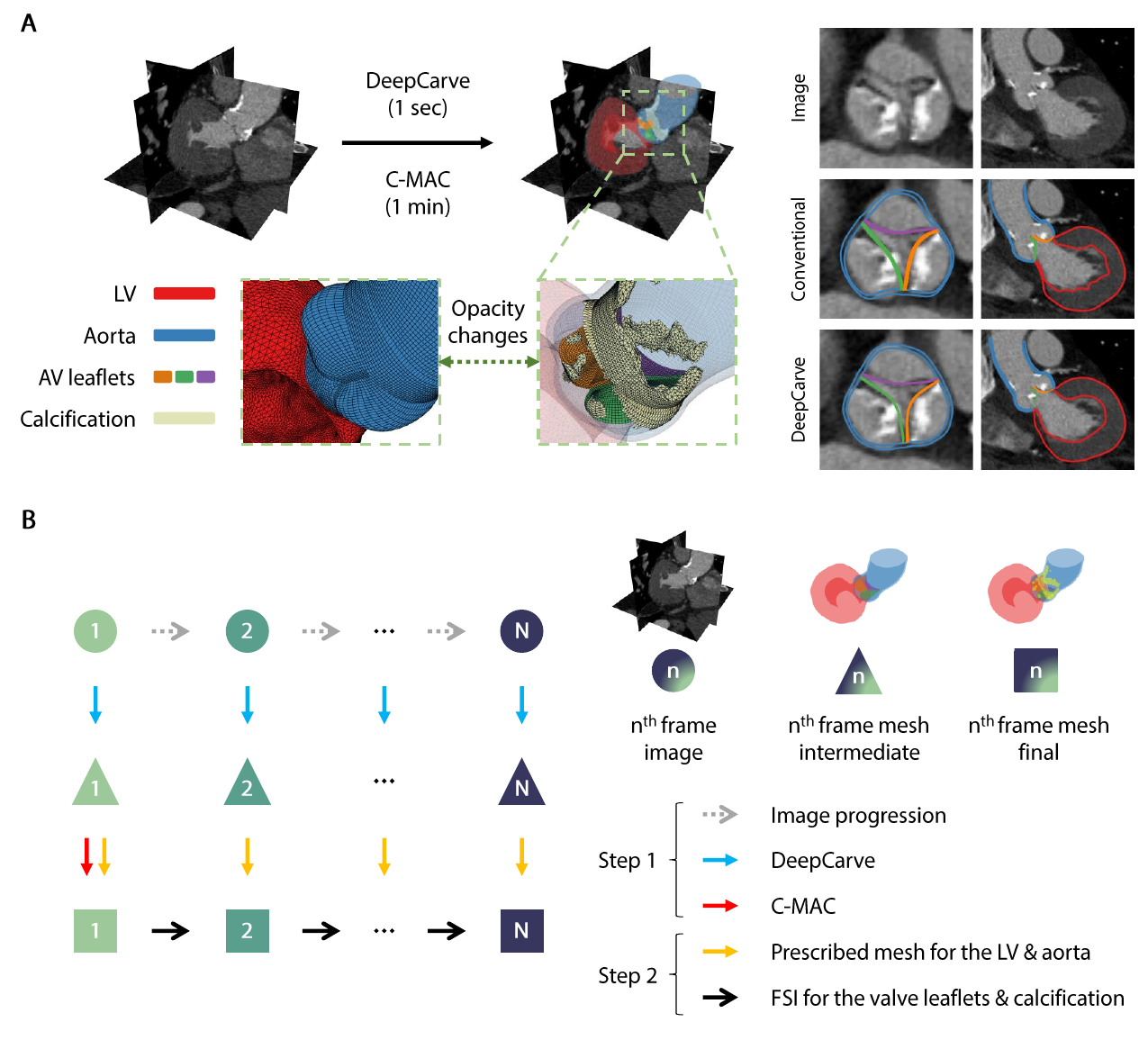}}
   {\bf Fig. 2. }{\textbf{Overview of our AI-powered geometry reconstruction algorithms for computational simulations.} \textbf{(A)} Visualization of our AI-generated simulation meshes' element quality and spatial accuracy. \textbf{(B)} Schematic of our multi-frame meshing approach for computational flow simulations.}
\end{figure}
Additionally, the modified DeepCarve algorithm substantially outperformed the original DeepCarve for the average LV surface accuracy (1.62 $\rightarrow$ 0.87), demonstrating the benefit of modeling the LV using surfaces instead of solids. The overall surface accuracy of the aorta and leaflets also improved (0.84 $\rightarrow$ 0.60), albeit at a slight cost to the volumetric element quality due to the division of the thickness layers in the valve leaflets. Nonetheless, no degenerate elements were produced by DeepCarve, as indicated by $\mid$ Jac $\mid \; <0$ being 0. This marks the first successful demonstration of the simultaneous surface and volumetric meshing of the AS geometry and the capability of DeepCarve to accurately model extremely thin tri-layer hexahedral leaflet elements.
\begin{table}[t]
{\textbf{ Table 1. Spatial accuracy and element quality of the meshes generated by the baseline method and the modified DeepCarve.} CD: normalized symmetric chamfer distance, $\mid $ Jac $ \mid$: scaled jacobian determinant. Values are mean $\pm$ std across all test-set patients, except for $\mid $ Jac $ \mid \; < 0$, which is the sum. $^*$ indicates $p<0.05$ between the two rows.}

\label{table:results_ca2_post_process}

\def\sym#1{\ifmmode^{#1}\else\(^{#1}\)\fi}
\sisetup{detect-weight,mode=text}
\robustify\bfseries

\begin{tabular*}{\linewidth}{ l @{\extracolsep{\fill}} *{2}{S[table-format=1.2(2),separate-uncertainty=true,table-align-text-post=false]} c *{2}{S[table-format=1.2(2),separate-uncertainty=true,table-align-text-post=false]}}

\toprule

& {\bfseries\makecell{CD (mm) $\downarrow$ \\ LV}} & {\bfseries\makecell{CD (mm) $\downarrow$ \\ Aorta \& AV}} & {\bfseries\makecell{$\mid$ Jac $\mid \; < 0$  $\downarrow$\\ Aorta \& AV}} & {\bfseries\makecell{$\mid$ Jac $\mid$ $\uparrow$ \\ Aorta \& AV}} & {\bfseries\makecell{Skew $\downarrow$ \\ Aorta \& AV}} \\

\midrule

\makecell{Conventional \\ registration}  &           0.93 \pm 0.13    &           0.84 \pm 0.19    &           14 &           0.71 \pm 0.04    &           0.52 \pm 0.05    \\

\midrule[0pt]

\makecell{DeepCarve}              &  0.87 \pm 0.12$^*$ &  0.60 \pm 0.17$^*$ &   0 &  0.85 \pm 0.02$^*$ &  0.36 \pm 0.03$^*$ \\

% \midrule[0pt]

\bottomrule

\end{tabular*}

\end{table}
Following DeepCarve, we applied C-MAC without modification to automatically incorporate calcification meshes while maintaining the existing heart mesh topology. The qualitative visualization demonstrates accurate, high-quality meshes for all components of the calcified AV and LV (Fig.~2A). The robustness of the final meshes were also verified by the successful completion of downstream analyses without any mesh degeneracy errors.

We performed dynamic computational modeling of AS hemodynamics using two steps (Fig.~2B): DeepCarve for the time-resolved geometry reconstruction and fluid-structure interaction (FSI) simulations for the detailed valve leaflet dynamics. For each patient, we applied the modified DeepCarve to each of the 8-10 CT phases independently to obtain the corresponding time-resolved meshes. Owing to our deep learning-based template deformation strategy and the shared imaging features between each time frame, the independently predicted meshes exhibited excellent intra-patient mesh correspondence, similar to \cite{kong2022learning}. Thus, we used the resulting meshes directly for motion interpolation during simulations. The total processing time of DeepCarve for all cardiac phases was around 2 seconds on an NVIDIA GPU RTX3080Ti laptop workstation, a 100-fold speed increase from conventional registration approaches. C-MAC was only applied to the first-frame mesh, as the calcification meshes for all other frames are positioned based on the motion of the attached structural components.
% The details of the FSI formulation and the simulation results will be discussed in later sections.

% we generated the initial flow into the aortic valve with image-derived LV motion. The overall modeling process involves 
% The details of the motion interpolation and leaflet simulations will be described in later sections.

% \begin{figure}[!t]
%     \centerline{\includegraphics[width=1\textwidth,trim=0 0 0 0 ,clip]{figures/Fig4.pdf}}
%    {\bf Fig. 4. }{Automated 3D printing mesh conversion algorithms for the (A) calcification, (C) valve leaflets, and (D) the combined LV and aorta.}
%     \label{fig:mit_flow_bar_plots}
% \end{figure}

% \cag{Might be good to separate 3D printing part and give a new subheading?}

\subsection*{Robust automated conversion to 3D printing geometry}

Our framework facilitates straightforward repurposing of the auto-generated meshes for 3D printed benchtop models. The differences in the meshing criteria necessitate conversion algorithms for the 3D printing geometry, but the entire process can be simplified to a single-button click with a total processing time of 10 seconds for each frame (Fig.~3A), due to the strong inter-part relationships established by our AI-powered mesh reconstruction methods. The component-agnostic meshing criteria for 3D printing include (1) using only triangular elements, (2) generating more uniform and smoother elements, and (3) ensuring overlapping surfaces for multi-part attachments. The head-to-head comparison between the AI-generated simulation mesh and its derivative 3D printing mesh demonstrates that our conversion algorithms can easily satisfy all desired criteria while maintaining great spatial consistency to the original mesh (Fig.~3A).

\begin{figure}[hbt!]
    \centerline{\includegraphics[width=1.0\textwidth,trim=0 0 0 0 ,clip]{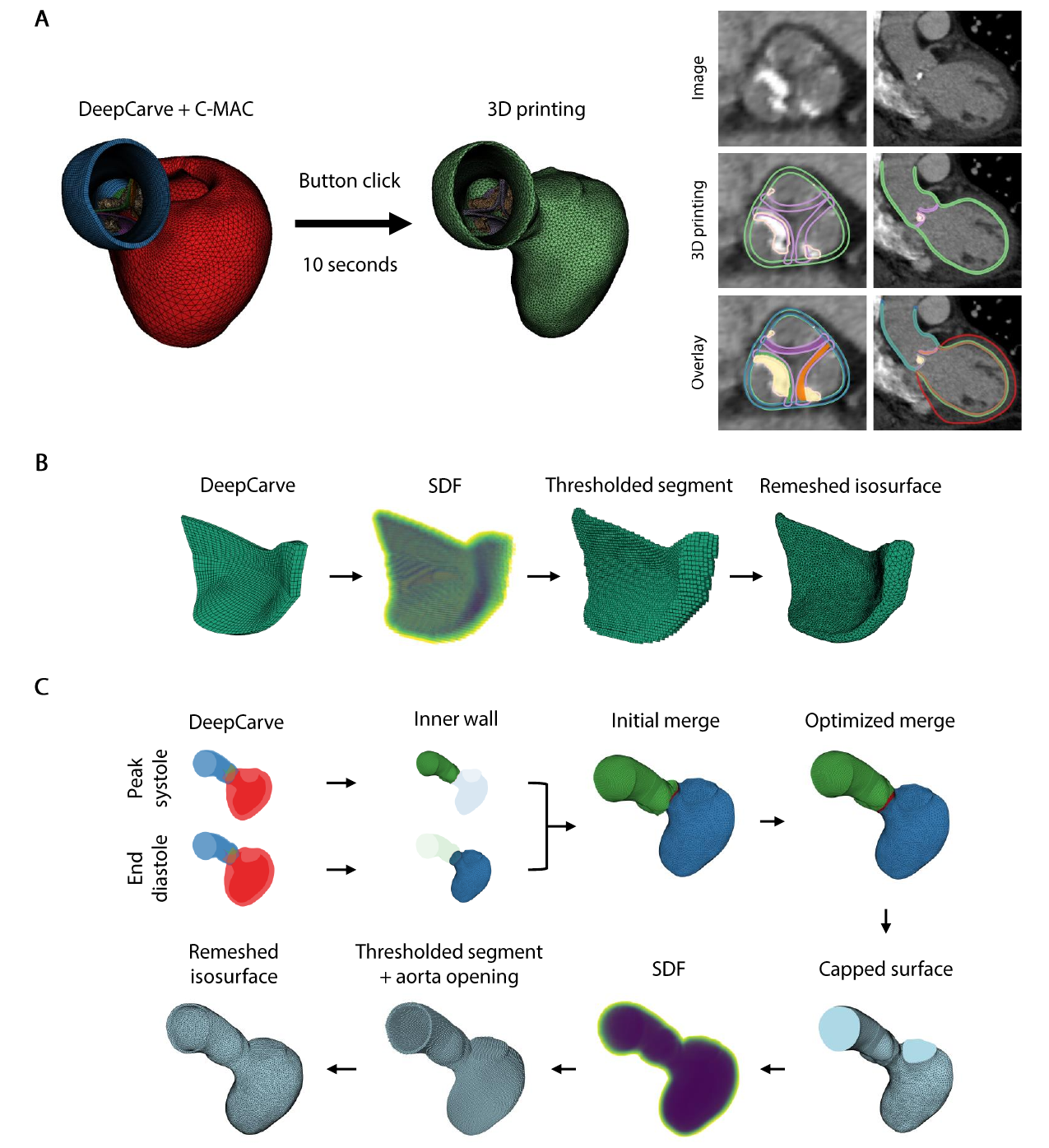}}
    {\bf Fig. 3. }{\textbf{Robust automated conversion to 3D printing geometry.} \textbf{(A)} Qualitative demonstration of the quality of the 3D printing mesh and its spatial consistency to the corresponding simulation mesh. Detailed steps of the conversion algorithms for the \textbf{(B)} AV leaflets and \textbf{(C)} the combined LV and aorta.}
    \label{fig:mit_results_3d_print}
\end{figure} 
% This underscores a key strength of our top-down approach; given a volumetric simulation mesh, conversions to simpler geometric representations can be easily automated.

% We satisfied all of the meshing criteria  by utilizing conversions to and from voxelgrid representations, as well as robust geometric manipulations within the voxelgrid domain.

Three structural components are often considered for AS benchtop modeling: calcification, valve leaflets, and the inner wall of the aorta and LV \cite{hosny2019pre,rosalia2023soft, kovarovic2021patient}. We achieved calcification conversion by extracting the isosurface from C-MAC's post-processed segmentation and subsequently applying standard remeshing and smoothing operations. We performed leaflet conversion with additional thickness adjustment steps, where we used finer-resolution voxelgrid signed distance functions (SDF) to robustly extract surfaces at various level sets (Fig.~3B). These steps enabled robust automated tuning of the leaflet thickness during 3D printing iterations, which helped establish compatibility with the 3D printer resolution and satisfy the surface overlap criteria for multi-part attachments.

To obtain a single-frame 3D printing mesh of the aorta and LV, we can simply apply the same algorithm as the leaflets while using the closed inner wall surface as the SDF boundary. However, to further tailor the model for soft robotics applications, we incorporated additional steps to combine the heart geometries from two different cardiac phases (Fig.~3C). We utilized the peak systolic AV geometry to model the valve's open state, adapting to the limitations of standard 3D printers in replicating the soft leaflet material properties for dynamic movements. We used the end diastolic LV to model the most relaxed LV state, from which the LV motion throughout the entire cardiac cycle can be reproduced via soft robotic contractions. To address the apparent position mismatch from the initial merging of the two time-separated meshes, we performed mesh node displacement optimizations to generate a smooth inner wall surface, which then served as an input to the remaining meshing steps resembling leaflet conversion (Movie S3).

The combination of geometries at two time points was designed specifically for our experimental capabilities, but different combinations of our conversion algorithms can generate a wide variety of 3D printable meshes for physical models. The robust automated conversion was made possible by the strong inter-part relationships established by our AI-powered mesh reconstruction methods.

\begin{figure}[hbt!]
    \centerline{\includegraphics[width=0.74\textwidth,trim=0 0 0 0 ,clip]{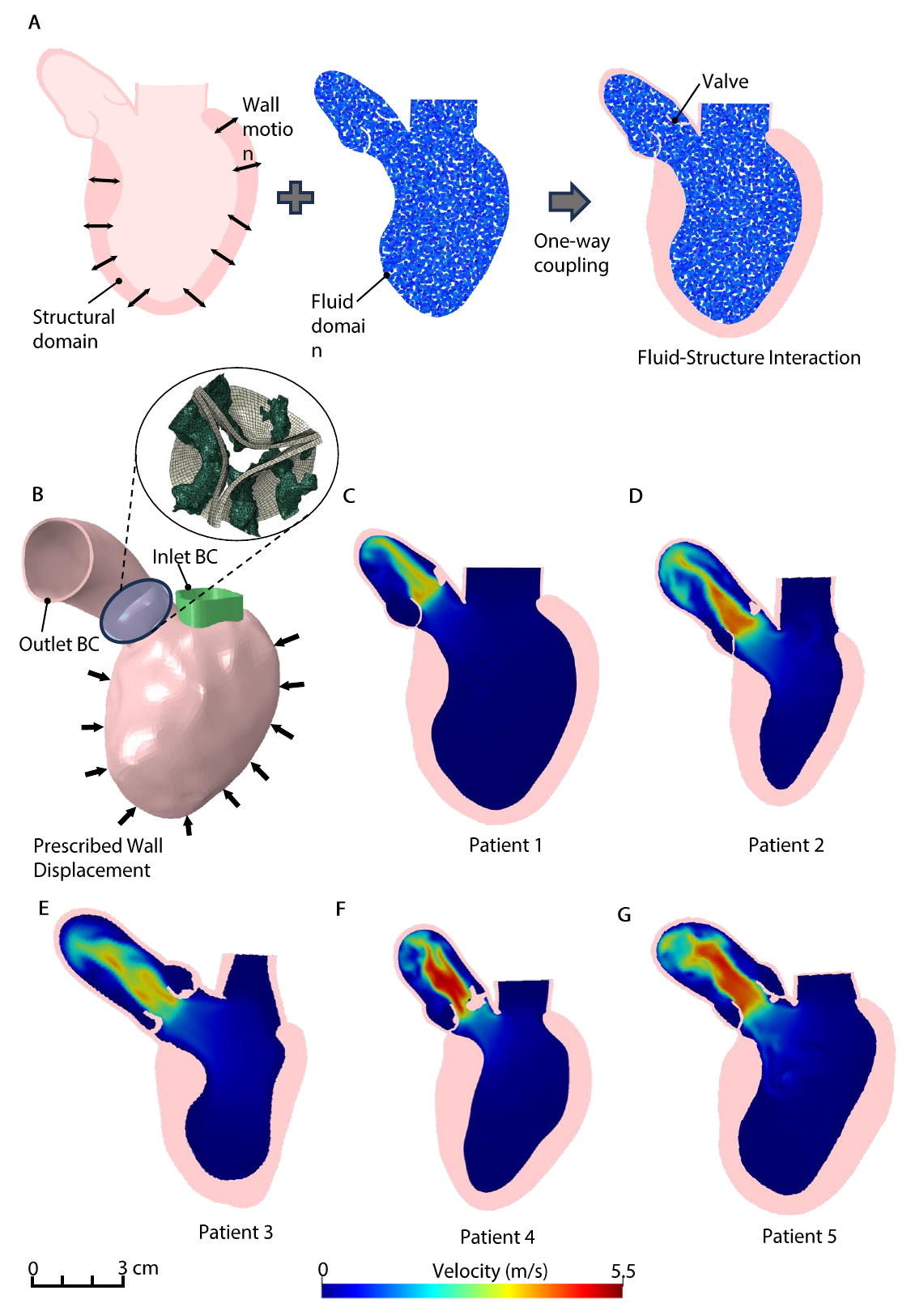}}
    {\bf Fig. 4. }{\textbf{ Overview of the computational model and patient-specific flow patterns.} \textbf{(A)} Schematic of FSI approach \textbf{(B)} Illustration of AI-powered simulation ready mesh. Schematic of one-way coupling, nodal displacement prescribed to LV wall. \textbf{(C-G)} Velocity map distributions in the anterior-posterior plane for patients 1 to 5, respectively. Velocity fields are captured during the systolic temporal frame when the AV velocity is at its peak state in the cardiac cycle. }
    
\end{figure}
\subsection*{Computational fluid model and insilico hemodynamic assessment }

We developed a computational framework that includes a FE model of the heart using AI-generated meshes of LV, aorta, AV, and calcium (Methods). Coupling the FE model of the heart with computational fluid dynamics (CFD) utilizing FSI architecture allowed us to simulate the global behavior of the patient-specific left side heart and hemodynamics (Fig. 4A). The DeepCarve algorithm was used to reconstruct time-resolved geometries, which were then utilized for motion interpolation of the heart (Movie S4). We used the first-frame mesh from DeepCarve and C-MAC to place the valve and calcification meshes. By prescribing nodal displacement to the LV mesh, we were able to simulate heart motion based on dynamic CT data and assess patient-specific hemodynamics (Fig. 4B). Using a comparison of global flow patterns and velocity magnitudes, we were able to qualitatively assess the AS hemodynamics specific to each patient. Figure 4C-G shows the cross-sectional velocity contours for five patient heart models during the peak systole. The stenotic morphology of the valve leaflets caused the flow jets to be more prominent at the contraction of the heart (Movie S4). Despite variations in heart size and boundary conditions among patients, the maximum velocity consistently occurred at the valve orifice in each patient/disease scenario, as expected. Following a remarkable elevation of the velocity in proximity to the valve orifice, the flow demonstrated a plug-like flow profile, remaining relatively stable within the jet orifice. The peak AV velocity was determined when the flow velocity across the AV reached its peak at the tips of the cusps. Patient 4 reached a peak velocity of 5.4 m/s, whereas Patient 1 had the lowest peak velocity of 4 m/s.

\subsection*{Recreating patient-specific valve anatomies and AS hemodynamics}

To assess the efficacy of our computational framework in obtaining accurate hemodynamic data, we conducted a comparative analysis with the conventional algorithm (Methods) (Fig. 5A). To re-create clinical measurements obtained via ultrasound imaging, we calculated the velocity profile across the AV by placing a short axis plane above the valve orifice (Fig. 5B). The cutting plane denotes the monitoring location for velocity measurements through the valve. Patients' clinical echocardiography data served as our ground truth when comparing hemodynamic findings from the computational models. For quantitative analysis, a set of global hemodynamic metrics, commonly used in clinical settings, was calculated from the FSI simulations with both the DeepCarve model and the conventional model (CM) (Fig. 5D-G, and Fig. S4). From the simulation, we determined the velocity values (V\textsubscript{peak}; Fig. 5D) and the effective orifice area (EOA) of the AV by calculating the opening area of its deformed shape at near peak systole (Fig. 5C). Likewise, we calculated the peak (dP\textsubscript{max}; Fig. 5E) and mean (dP\textsubscript{mean}; Fig. 5F) transvalvular pressure gradient by measuring the flow upstream and downstream through planes above and below the valve annuli. Figure 5G reports another clinical metric, left ventricular ejection fraction (LVEF), based on the volumetric change of LV through the cardiac cycle. The results show that the FSI simulations and the subject-specific echo data follow the same trends and are of comparable magnitude. The simulations using the DeepCarve model more closely matched the clinical echo data in all target metrics, indicating a mean absolute deviation of 3.8 ± 1.3\% for all metrics compared with the conventional approach (12.4 ± 3.4\%). We calculated the deviations in V\textsubscript{peak} to be 2.4 ± 1.5\% compared to 9.8 ± 1.6\%, dP\textsubscript{max} as 5.3 ± 0.9\% compared to 13.9 ± 3.1\%, dP\textsubscript{mean} as 4.1 ± 1.3\% compared to 14.0 ± 4.1\%, and LVEF as 3.3 ± 2.9\% compared to 12.1 ± 8.2\% for DeepCarve and conventional methods. 
\begin{figure}[hbt!]
    \centerline{\includegraphics[width=0.67\textwidth,trim=0 0 0 0 ,clip]{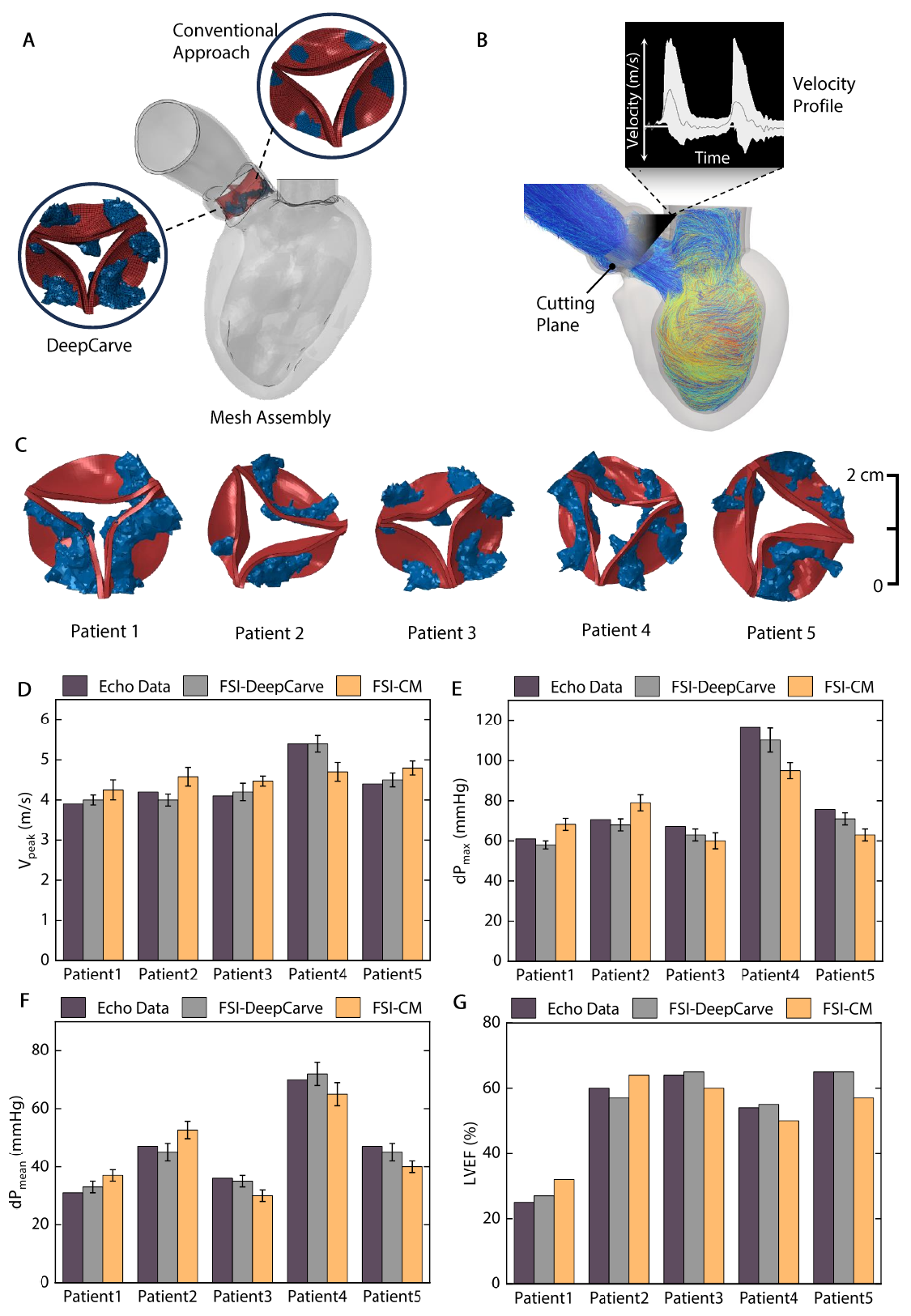}}
    {\bf Fig. 5. }{\textbf{Comparison of patient-specific hemodynamic findings with AI-powered and conventional algorithms.} \textbf{(A)} Constructed valve and calcium geometry using DeepCarve and conventional meshing. \textbf{(B)} The visualization of particle pathlines based on   FSI simulations. The cutting plane denote the probe location for AV velocity measurements. \textbf{(C)} AV profiles of patients 1-5 during peak systole in the cardiac cycle. Comparison of FSI simulations with DeepCarve and conventional meshing/registration, shown by measured hemodynamics metrics of \textbf{(D)} V\textsubscript{peak}, \textbf{(E)} dP\textsubscript{max}, \textbf{(F)} dP\textsubscript{mean}, \textbf{(G)} LVEF, demonstrating that DeepCarve-FSI allows for more accurate hemodynamic measurements and agrees well with the ground truth (clinical echo data). Each error bar represents means ± 1SD (n= 3 consecutive heart cycles).}
    \label{fig:mit_flow_bar_plots}

\end{figure}
\subsection*{3D printing patient-specific meshes and in vitro hemodynamic assessment using echocardiography}
We adopted our mock circulatory left heart flow loop to test the functionality of our automated mesh conversion algorithm for 3D printing and benchtop testing \cite{rosalia2023soft}. The soft robotic sleeve served as the actuator for the 3D-printed heart and drives fluid flow in the circuit, without the need for an additional pump (Fig. 6A). The LV sleeve was tailored to fit the patient's anatomy and activated to mimic anatomical filling, emptying, and wall motion during systole. To validate the computational findings and the accuracy of our mesh conversion framework for 3D printing, we manufactured the anatomies of Patient 1 and 2 (Fig. 6B), and recreated their hemodynamic profiles in our in vitro system (Fig. S5). The automated mesh conversion process successfully met the watertightness requirement and did not pose any surface overlap or compatibility issues with the 3D printer. We primarily used echocardiography to compute the flow velocity through the AV and at the left ventricular outflow tract (LVOT) (Movie S5). The representative continuous Doppler waveforms in Figure 6C showcase the aortic velocity profiles for Patients 1 and 2. The measured metrics of V\textsubscript{peak},V\textsubscript{LVOT}, dP\textsubscript{mean}, dP\textsubscript{max}, and LVEF closely matched with the clinical patient data in both cases. For each of these metrics, we computed the compounded absolute deviation (\%) from the corresponding clinical echo report and presented the similarity between the FSI and the in vitro studies (Fig. 6D). The results indicated that the FSI simulations using conventional methodologies had the highest deviation of 11.1 ± 3.2\%, while the DeepCarve approach exhibited the lowest deviation of 5.9 ± 1.1\% compared to the in vitro value of 8.8 ± 2.0\%. These findings suggest DeepCarve-FSI approach agrees more closely with the clinical echo data, demonstrating the accurate performance of our AI-powered computational framework. Analogously, the in vitro test results exhibited similar patterns as the clinical data but had a comparatively higher margin of error than the DeepCarve-FSI technique, which will be discussed later. The deviation in LVEF in in vitro studies was slightly less prominent than other metrics and was within the same ranges of clinical data and DeepCarve-FSI. This is likely due to our automated mesh conversion for 3D printing, which results in accurate LV volumes at both end-systole (ESV) and end-diastole (EDV). Overall, these data showcase our platform's capability to replicate LV hemodynamics accurately and quickly in a benchtop setting.

\begin{figure}[hbt!]
    \centerline{\includegraphics[width=0.85\textwidth,trim=0 6cm 0 0 ,clip]{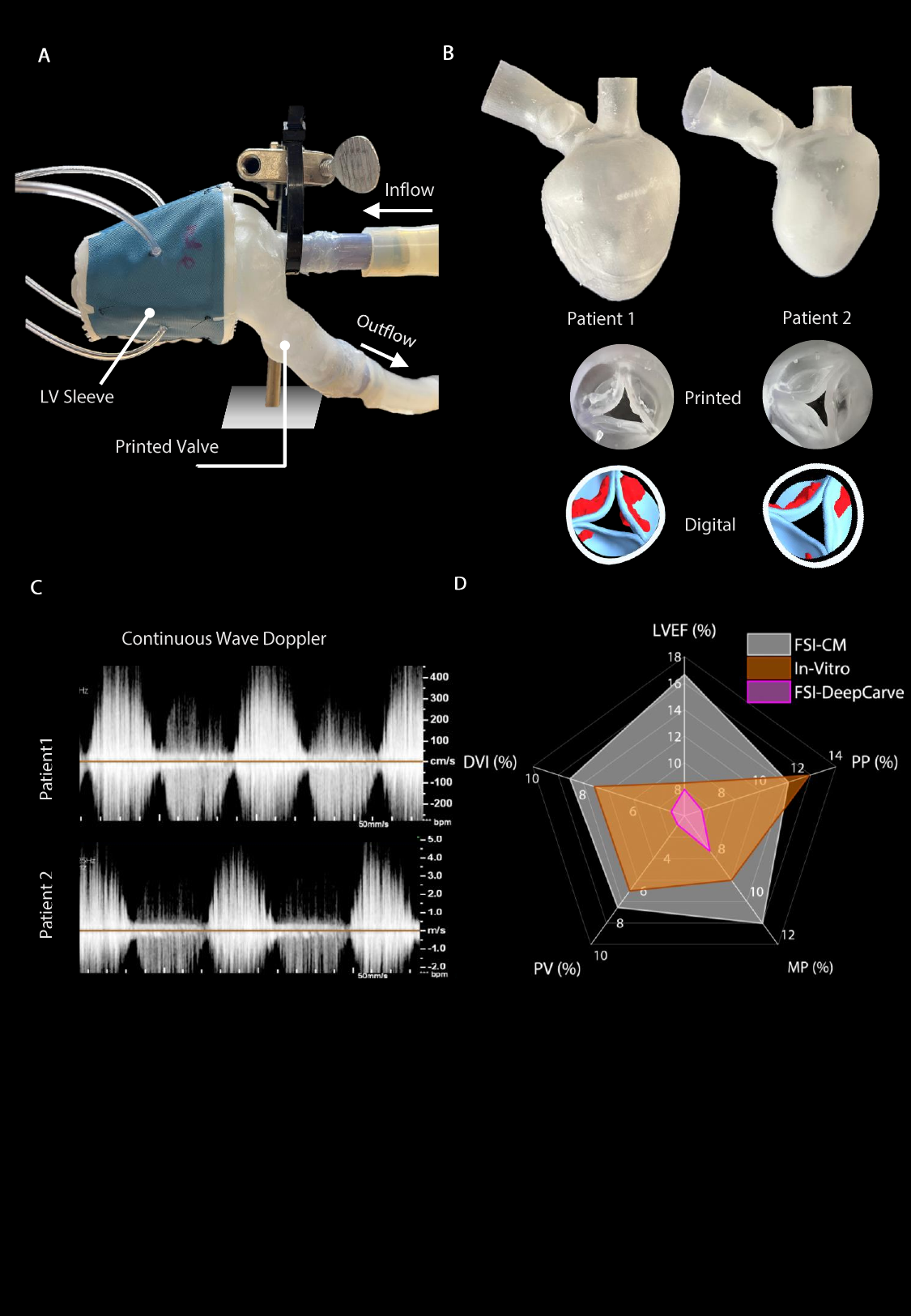}}
    {\bf Fig. 6. }\textbf{A platform for testing AI-powered 3D printing mesh with soft robotic models. } \textbf{(A)} Overview of soft robotic system. \textbf{(B)} As an example, the geometries of Patients 1 and 2 were 3D printed using our automated conversion. \textbf{(C)} Representative echocardiographic velocity waveforms across the aortic valve from CW Doppler measurements during actuation by the soft robotic LV sleeve in three consecutive cardiac cycles. \textbf{(D)}  The similarities between clinical metrics estimated from in vitro experiments and in silico simulations with FSI using conventional meshing/registration method (FSI-CM) and DeepCarve (FSI-DeepCarve). 
    Clinical parameters include LVEF, mean pressure gradient (MP), peak pressure gradient (PP), dimensionless velocity index (DVI), and peak aortic velocity (PV).

\end{figure}

\section*{Discussion} 

In this paper, we propose and validate a start-to-end computational pipeline for patient-specific hemodynamic analysis for AS using CT imaging patient data. We extend the DeepCarve method to a multi-modal tool that generates fully coupled patient-specific FSI and physical benchtop models of the left heart. Compared to traditional techniques, the proposed method demonstrates improved geometry reconstruction at much faster test-time performance. The reconstruction quality is particularly important for accurately representing the highly dynamic and thin AV leaflets and irregularly dispersed calcium deposits, both of which are crucial for simulating AS. Our solution entails using DeepCarve on each individual CT time frame that exploits the deep learning template deformation strategy on shared image features across time frames. Additionally, we showcase the direct and automatic transformation of our AI-generated meshes into meshes suitable for 3D printing. Our robust volumetric mesh outputs exhibit well-defined multi-part relationships between different heart components, which allows our automated transformations to easily meet the necessary meshing criteria while maintaining high/elevated spatial consistency to the simulation mesh. 

We used clinical data from a cohort of AS patients to validate our approach as a platform for prediction of hemodynamic diagnostic metrics of AS. Our model was shown to generate the left heart kinematics as visualized by CT data and predict hemodynamic changes associated with patients' disease. As a result, our AI-powered FSI model not only recapitulates the diverse range of clinical metrics of AS, but also offers accurate and efficient solution in comparison to traditional methods.

Some limitations of this study are noted. The primary shortcoming of our meshing approach is the limited flexibility of the template deformation-based approach. For complex anatomies or, for example, congenital heart defects, we would need a modified template for each phenotype. Our FSI method could predict AS hemodynamics despite the intrinsic limitation of current imaging techniques. However, it was based on several assumptions. For example, we assumed a healthy leaflet condition and assigned the same hyperelastic material properties to the valve leaflets regardless of the individual patient's disease state (Methods). Similarly, we implemented the same linear elastic material properties for calcium geometries. Another limitation is the relatively lower resolution (both spatial and temporal) of the original CT data at which the valve leaflet dynamics are captured by the data-driven meshing approach. Although our model can effectively create the dynamics of the valve leaflet even with low resolution of CT data, the data-driven meshing technique necessitates a fully coupled analysis to simulate the valve's behavior during an ejection period, leading to longer simulation periods. It is also worth noting that our computational FSI framework uses a commercial solver package (Methods) and requires some manual operations to integrate the mesh outputs from DeepCarve into the simulation platform. To fully automate the planning workflow from a raw CT scan to personalized hemodynamic results, an open-source platform for FSI modeling would be necessary. The \textit{in vitro} testing framework also contains some limitations. Our 3D printing method allows for the production of soft and flexible structures, but its material characteristics still result in an excessive hypokinetic state of the valve leaflet due to limitations in the 3D printing techniques that we adopted, which required a minimum thickness of 0.6mm of the leaflets. While it may seem rational to use fixed leaflets to calculate the flow profile of AS during peak systole, there is a potential mismatch in accurately determining the effective area opening during this phase. This is due to the limited temporal resolution of the original CT data, which offers only eight time frames per cycle. 

Further work might enhance the accuracy of the valve dynamics through AI to tackle the constraints posed by low-resolution CT data and also eliminate the need for any specific material definitions for leaflets and calcium. This approach would further reduce the simulation time while replicating not only the heart but also the precise valve motion detected in the original CT data. Some of the limitations could be also addressed in the future with higher-resolution CT datasets. By using super-resolution CT images, we can identify the valve dynamics and unique stenotic profile more accurately. This can potentially speed up the translation to clinical use by reducing the overall time required for the computational pipeline from start to end. Additionally, it would enable the creation of accurate 3D printed models that replicate the hemodynamics of AS even more accurately. 
Alternatively, for developing efficient in vitro platforms, we could seek advancements in multi-material 3D printing strategies that might allow us to manufacture enhanced valve leaflets and calcium with accurate anatomical thicknesses, and biomimetic mechanical properties. In future work, we aim to expand the range of clinical scenarios that we can simulate, beyond AS. We seek to focus on interventional and surgical approaches for AS by recapitulating the correct hemodynamics specific to each patient. For instance, our computational pipeline may assist in planning TAVR surgery and identifying subgroups within AS populations where TAVR might be advantageous and delivered without any risk. 

This work addresses two major but often overlooked challenges in developing preclinical cardiovascular models (\textit{in silico}, \textit{in vitro}) for patients with AS. First, current image segmentation protocols rely heavily on manual work and pose significant inter and intra-expert variability. The proposed technique significantly reduces the computational time required for segmentation and meshing, from several hours of manual work to approximately 1 minute of automated computing. Additionally, it provides higher accuracy of patient-specific heart, valve, and calcium mesh. Second, the complexity and high computational costs associated with multiphysics heart models, along with the requirement for skilled researchers to operate them, present a major barrier to the translation of digital (\textit{in silico}) and benchtop (\textit{in vitro}) twins in clinical practice. The process of creating patient-specific 3D geometry, such as preparing a mesh suitable for simulation, separating the entire body into multiple, contact-free parts, determining heart motion, and assigning electromechanical tissue properties and appropriate boundary conditions to achieve a beating motion, collectively demands substantial computational resources and effort from trained researchers. Similarly, the accuracy of mimicking heart motion and AS flow profile in benchtop models relies heavily on the quality of image segmentation and 3D printed geometry. Any discrepancy between the patients' CT images and 3D printed anatomies compromises this accuracy, thereby reducing the feasibility of using these hydrodynamic models in a clinical setting. The proposed FSI pipeline with DeepCarve addresses these issues by recapitulating the patient-specific heart motion and AS hemodynamics without requiring any sophisticated electromechanical heart models, massive supercomputers, or days-long simulation time. 
In conclusion, the computational pipeline presented in this work offers fast, high-fidelity simulation capabilities and has potential to accelerate the development of intracardiac interventions and tools for evaluating and treating AS. Our DeepCarve-FSI technique delivers an effective solution for reproducing the left heart dynamics in both computational and benchtop models. It could be adapted to use in clinics for device selection for TAVR, procedural planning, and outcome prediction. We are hopeful that our model may further advance cardiac twin models by reducing the computational complexity involved in modeling AS and other cardiovascular conditions.

\section*{Methods}

\subsection*{Data acquisition and pre-processing}

A total of 79 3D CT scans were collected to train and geometrically evaluate the DeepCarve algorithm. 65 pre-TAVR CT scans were obtained from 54 patients at Hartford hospital with an IRB-approved protocol. The remaining 14 scans were from the training set of the MM-WHS public dataset, which mostly consisted of volunteer subjects without AS \cite{zhuang2016multi}. The testing set was ensured to never have any overlapping patients with the training and validation sets. The training/validation/testing splits were 35/9/35. All scans showed the entire LV myocardium and tri-leaflet AV with different levels of calcification. The raw image in plane pixel spacing was in the range [0.281, 0.727] mm, with a median of 0.488 mm, and the slice thickness was in the range [0.450, 1.000] mm, with a median of 0.625 mm. Slices were acquired in the axial view.

The associated ground-truth labels included the surface meshes of (1) the inner wall of the ascending aorta, (2) the inflow side of the AV leaflets, and (3) the LV myocardium. The labels were obtained using semi-automated methods as outlined in \cite{pak2023patient}. The CT intensity was normalized by linearly mapping [-158, 864] Hounsefield Units (HU) to [0, 1], and the voxel spacings were resampled to an isotropic resolution of 1.25mm with a fixed field of view of 128 voxels in each axis.

For hemodynamics modeling, 5 additional time-resolved pre-procedural CT scans were collected from 5 different TAVR patients at Hartford hospital. The imaging characteristics and pre-processing steps were identical to above. For clinical validation, the corresponding pre-TAVR reports from comprehensive echocardiography were used.

\subsection*{Modified DeepCarve training details}

The modified DeepCarve in this work utilizes (1) the surface representation of the entire LV myocardium and (2) the tri-layer hexahedral elements of the AV leaflets. These changes were implemented by first modifying the original template using Solidworks (Dassault Systèmes, 2023) and Hypermesh (Altair, 2022), and then modifying the training loss to be compatible with the simultaneous optimization of surface and volumetric meshes. In addition to the new templates and training loss, a random translation augmentation was added to the original training procedure, similar to C-MAC's implementation of the model \cite{pak2024robust}. The training set-up was otherwise identical to DeepCarve \cite{pak2023patient}. Briefly, a U-net architecture was employed with extra skip connections and a modified prediction head to predict b-spline diffeomorphic deformation fields from 3D CT. The exact architectural definitions can be found in \cite{pak2021weakly,pak2021distortion}. We used the Adam optimizer \cite{kingma2014adam} with a fixed learning rate of 1e-4, batch size of 1, and 8000 training epochs.

The training loss function for our modified DeepCarve was
\begin{equation} \label{eq:overall_loss}
    \mathcal{L}
    = \Bigg[ \frac{1}{N} \sum_{n=1}^N \big[ \mathcal{L}_{ch} \big]_n \Bigg] + \underbrace{\lambda_1 \mathcal{L}_{ARAP} + \lambda_2 \mathcal{L}_{ASqrt}}_{\mathcal{L}_{solid}} + \underbrace{\lambda_3 \mathcal{L}_{normal} + \lambda_4 \mathcal{L}_{lap} + \lambda_5 \mathcal{L}_{edge}}_{\mathcal{L}_{surf}}
\end{equation}

\noindent where $n$ denotes the mesh component index, and $N=5$ for the LV myocardium, aorta, and 3 aortic leaflets. $\lambda_i = \{1, 10, 10, 10, 1\}$ for each loss component, respectively, via hyperparameter optimization. Due to the template element characteristics, $\mathcal{L}_{solid}$ entirely governed the element quality of the aorta and the valve leaflets, whereas $\mathcal{L}_{surf}$ entirely governed the element quality of the LV surface.

The chamfer distance was used to optimize for surface accuracy \cite{wang2018pixel2mesh}
\begin{equation}
    \mathcal{L}_{ch} = \frac{1}{\mid A \mid}  \sum_{\mathbf{a} \in A} \min_{\mathbf{b} \in B} \norm{\mathbf{a}-\mathbf{b}}_2^2 + \frac{1}{\mid B \mid} \sum_{\mathbf{b} \in B} \min_{\mathbf{a} \in A} \norm{\mathbf{b}-\mathbf{a}}_2^2
\end{equation}

\noindent where $A$ and $B$ are two sets of points and $\mathbf{a}$ and $\mathbf{b}$ are the set elements, each consisting of 3D point coordinates. The chamfer distance was calculated separately for each component (denoted as $[\mathcal{L}_{ch}]_n$), and then averaged across different components to obtain the final spatial accuracy metric.

To ensure high-quality solid elements, a combination of isotropic and anisotropic distortion energies was used, similar to DeepCarve \cite{pak2023patient}. The as-rigid-as-possible (ARAP) energy was used for the isotropic component
\begin{equation}
\mathcal{L}_{ARAP} = \frac{1}{K} \sum_{k=1}^{K} \norm{\mathbf{F}_k - \mathbf{R}_k}_F^2
\end{equation}

\noindent where $k$ denotes the solid element index, $K$ is the total number of solid elements, and $\mathbf{F}_k$ and $\mathbf{R}_k$ are the deformation gradient and the rotational component of the deformation gradient for the $k^{\textrm{th}}$ element. The definitions of $\mathbf{F}$ and $\mathbf{R}$ based on the element geometry can be found in \cite{kim2020dynamic,pak2023patient}. This term represents the uniformly weighted average of the isotropic ARAP energy densities of all solid elements.

The ASqrt energy, a square root variation of the popular anisotrpic St. Venant Kirchoff energy, was used for the anistropic energy component \cite{kim2020dynamic}
\begin{equation}
\mathcal{L}_{ASqrt} =  \frac{1}{K} \sum_{k=1}^{K} \Big(\sqrt{\mathbf{d}_k^T \mathbf{F}_k^T \mathbf{F}_k \mathbf{d}_k} - 1 \Big)^2
\end{equation}

\noindent where $\mathbf{d}_k$ denotes the anisotropy direction for calculating the energy density for the $k^{\textrm{th}}$ element. Similar to $\mathcal{L}_{ARAP}$, the anistropic energy densities were averaged uniformly across all solid elements. Following DeepCarve, $\mathbf{d}_k$ was set as the thickness direction of each hexhedral element.

Similar to previous works \cite{wang2018pixel2mesh,gkioxari2019mesh,pak2021weakly}, three surface element quality metrics were optimized for high-quality surface elements: surface normal consistency, Laplacian smoothness, and edge length correspondence. The surface normal consistency loss is commonly defined as
\begin{equation}
\mathcal{L}_{normal} =\; \frac{1}{\mid \mathcal{N}_f \mid} \sum\nolimits_{(\mathbf{n}_{fi}, \mathbf{n}_{fj}) \in \mathcal{N}_f} 1 - \frac{<\mathbf{n}_{fi}, \mathbf{n}_{fj}>}{\norm{\mathbf{n}_{fi}}_2 \norm{\mathbf{n}_{fj}}_2}
\end{equation}

\noindent where $\mathbf{n}_{fi}$ is the surface normal at $i^{\textrm{th}}$ element, and $\mathcal{N}_f$ is the set of surface normals in all neighboring faces. This term facilitates neighboring faces to maintain similar surface normals.

The uniform Laplacian smoothing loss was defined as
\begin{equation}
\mathcal{L}_{lap} =\; \frac{1}{\mid V \mid} \sum_{\mathbf{v}_i \in V}  \norm{\frac{1}{\mid \mathcal{N}(\mathbf{v}_i) \mid} \sum_{\mathbf{v}_j \in \mathcal{N}(\mathbf{v}_i)} \mathbf{v}_i - \mathbf{v}_j}_2
\end{equation}

\noindent where $V$ is the set of all vertices in the surface mesh, $\mathbf{v}_i$ is the $i^{\textrm{th}}$ vertex coordinates, and $\mathcal{N}$ is the set of neighbors for a chosen vertex $\mathbf{v}_i$. This term encourages each node to be in close proximity to the average of its neighboring node positions.

Finally, the edge length correspondence loss was
\begin{equation}
\mathcal{L}_{edge} =\; \frac{1}{\mid \varepsilon \mid} \sum_{(\mathbf{v}_i, \mathbf{v}_j) \in \varepsilon} \Big( \frac{\norm{\mathbf{v}_i - \mathbf{v}_j}_2}{\max\limits_{(\mathbf{v}_i^\prime, \mathbf{v}_j^\prime) \in \varepsilon} \norm{\mathbf{v}_i^\prime - \mathbf{v}_j^\prime}_2} - \frac{\norm{\phi(\mathbf{v}_i) - \phi(\mathbf{v}_j))}_2}{\max\limits_{(\mathbf{v}_i^\prime, \mathbf{v}_j^\prime) \in \varepsilon} \norm{\phi(\mathbf{v}_i^\prime) - \phi(\mathbf{v}_j^\prime)}_2} \Big)^2
\end{equation}

\noindent where $\epsilon$ is the edges represented as a set of two vertex coordinates comprising each edge, and $\phi(\mathbf{v}_i)$ represents an arbitrary displacement of each vertex. This term encourages similar edge length ratios before and after the mesh deformation.

\subsection*{Conventional registration}

For fair comparisons with DeepCarve, the conventional registration algorithm was chosen to perform fully automated registration, potentially with an offline model training component to help identify the target structure. To that end, registration algorithms that require any amount of manual annotations at test-time were excluded. For the offline model training, the models were additionally restricted to use a similar level of structural information as DeepCarve, i.e. surface mesh or segmentation.

Under these constraints, we followed a common approach that splits the registration into two steps: (1) automated segmentation and (2) segmentation-based template registration \cite{astorino2012robust,tavakoli2013survey,pouch2015medially}. For the former, a deep learning model was trained to perform multilabel segmentation of the aorta, LV myocardium, and AV leaflets. The training loss was defined as the weighted sum of the Dice similarity coefficient (DSC) \cite{pak2020efficient}
\begin{equation}
    \mathcal{L}_{DSC} = \sum_{n=1}^{N} \alpha_n \Bigg[ \frac{2 \sum_{i} p_i g_i}{\sum_{i} p_i + \sum_{i} g_i} \Bigg]_n
\end{equation}

\noindent where $\alpha_n$ is the component-specific weight, and the bracketed entry is the DSC of the $n^{\textrm{th}}$ structural component. $p_i$ and $g_i$ are the predicted and ground-truth segmentations, respectively, at the $i^{\textrm{th}}$ voxel. For training the segmentation model, we chose uniform weighting of $\alpha_n$ to ensure equal amount of optimization for delineating all five structural components.

We maintained almost an identical training procedure to DeepCarve, such as the neural network architecture, training augmentation, and optimizer. For training and validation, we used the intermediate segmentation labels that we generated during the ground-truth surface labeling workflow.
% We confirmed that the trained model performed similarly to existing segmentation models with qualitative and quantitative evaluations \dpak{add average evaluation dice metrics}.

Then, using the multilabel segmentation output, component-aware registration was performed using the weighted sum of the DSC as the overall spatial accuracy objective. We first performed similarity transformation (rotation, translation, and anisotropic scaling), and subsequently performed nonrigid deformation to fit each component of the template mesh to the predicted segmentation, similar to \cite{papademetris2006bioimage,tavakoli2013survey}. Since no regularizations were needed for the similarity transformation, the similarity registration loss was defined as
\begin{equation}
    \mathcal{L}_{sim} = \mathcal{L}_{DSC} \;\;\textrm{with different}\; \alpha_n
\end{equation}

\noindent where $\alpha_n = \{5,1,1,1,1\}$ to allow the largest structure, the LV, to most heavily influence the similarity registration outcome. Separate Adam optimizers \cite{kingma2014adam} were used for the scaling, rotation, and scaling parameters, with learning rates of 1e-1, 1e-1, and 1, respectively.

For non-rigid registration, the deformation field was regularized using the popular bending energy \cite{rueckert1999nonrigid}, and additionally optimized for the mesh edge length loss to improve the overall mesh quality. The deformation field was constrained to be b-spline \cite{rueckert1999nonrigid,sandkuhler2018airlab} with isotropic control point spacing of 3 voxels and diffeomorphic \cite{dalca2018unsupervised,sandkuhler2018airlab}, similar to the deformation from DeepCarve. The final non-rigid registration loss was defined as
\begin{equation}
    \mathcal{L}_{non\textrm{-}rigid} = \mathcal{L}_{DSC} + \lambda_1 \mathcal{L}_{bending} + \lambda_2 \mathcal{L}_{edge}
\end{equation}

\noindent where the $\mathcal{L}_{DSC}$ was calculated with uniform $\alpha_n$ to ensure accurate matching of every component, and  $\lambda_i = \{1e-2, 1\}$ via hyperparameter optimization. $\mathcal{L}_{bending}$ is defined as the following:
\begin{equation}
\begin{aligned}
      \mathcal{L}_{bending} = \dfrac{1}{V} \int\limits_0^X \int\limits_0^Y \int\limits_0^Z \bigg[ & \left(\dfrac{\partial^2 \phi}{\partial x^2}\right)^2 + \left(\dfrac{\partial^2 \phi}{\partial y^2}\right)^2 + \left(\dfrac{\partial^2 \phi}{\partial z^2}\right)^2  \\
&+ 2\left(\dfrac{\partial^2 \phi}{\partial xy}\right)^2 + 2\left(\dfrac{\partial^2 \phi}{\partial xz}\right)^2 + 2\left(\dfrac{\partial^2 \phi}{\partial yz}\right)^2 \bigg] \; dx \; dy \; dz
\end{aligned}
\end{equation}

\noindent where $V$ is the number of voxels and $\phi$ is the voxelwise deformation field.  $\mathcal{L}_{edge}$ is identical to the definition in the DeepCarve training details section. The Adam optimizer \cite{kingma2014adam} with a learning rate of 1e-2 was used for the deformation field optimization.

The conventional registration method was implemented in-house, mostly due to the lack of compatibility of existing libraries with our specific task. We required the methods to simultaneously register five different potentially overlapping segmentations and maintain reasonable mesh qualities for all components, both of which were difficult to enforce in standard registration libraries. Our implementation was carefully optimized for performance, while following the standard approaches for the deformation and regularization formulations found in the literature and software libraries.

% Therefore, we implemented an in-house conventional registration algorithm combining ideas from the literature. Note that this was our effort to obtain the most accurate registration algorithm, as all other available options performed far worse than our in-house implementation for our desired workflow. 

% Unfortunately, no libraries supported registration using multilable segmentation, which was required to match the template mesh to both the valve leafletes and the segmented blood pool inside the aorta. Additionally, we could not find existing libraries that supported deformation regularization based on surface mesh quality metrics. This often led to weird surfaces especially for the LV, where large deformations were often required to match to the highly variant myocardial volumes of our patient group.

\subsection*{Conventional calcification modeling}

For conventional modeling, the calcification was implemented as element-specific assignment of stiffer material properties, similar to \cite{morganti2014simulation,loureiro2020biomechanical}. The desired calcification elements were selected using the post-processed segmentation of C-MAC, the same geometry that was used to perform the final calcification meshing. The calcification segmentation value was trilinearly interpolated at each node of the conventionally registered mesh, and any elements with at least one node with an interpolated nodal segmentation value of $>$0.5 were assigned stiffer material properties.

\subsection*{Geometric evaluation metrics and statistical analyses}

The normalized symmetric chamfer distance was used to evaluate the average surface accuracy of the reconstructed meshes. The mathematical definition is $\frac{\mathcal{L}_{ch}}{2}$, where the division normalizes the chamfer distance that is calculated twice in both directions (i.e. prediction $\rightarrow$ label and label $\rightarrow$ prediction). The scaled jacobian determinant and skew were obtained using the VTK library \cite{schroeder1998visualization}. The metrics were first calculated for each component for each patient, and then combined across component groups via the weighted average using the number of points/elements as weights. The mean, standard deviation, and statistical significance were calculated from the patient-specific component-grouped metrics. This evaluation method is identical to DeepCarve. For statistical analyses, paired t-tests with two-sided alternative hypothesis were performed using the SciPy package \cite{2020SciPy-NMeth}.

\subsection*{3D printing mesh conversion algorithms}

For all structural components, the final operations for 3D printing mesh conversion were (1) isosurface extraction from voxelgrid representations, (2) remeshing, and (3) mesh smoothing. The isosurface extraction and mesh smoothing were performed using VTK \cite{schroeder1998visualization} and remeshing using ACVD \cite{valette2008generic}. Since all of these automated operations can be robustly performed with reasonable input geometry and algorithmic parameters, the crux of the conversion process was obtaining an accurate voxelgrid representation for each component.

For calcification, the post-processed segmentation from C-MAC was used for the conversion. Since the post-processed segmentation was already at a high enough voxelgrid resolution of 0.33mm isotropic spacing to capture the contact surfaces, the extracted surfaces maintained great overlapping spatial relationships with its surrounding geometry.

For the leaflets, the surface of each leaflet hexahedral mesh was first extracted, and then converted to the corresponding segmentation using the image stencil operation. The segmentation was obtained at an isotropic 0.25mm voxel spacing to account for the thin structure of the leaflets and to allow for precise control over the final mesh thickness. The segmentation was converted to the SDF representation using SciPy's distance transform package \cite{2020SciPy-NMeth}, and then the final post-processed segmentation was obtained for meshing by variably thresholding the calculated voxelgrid distances. The thresholds were chosen mostly to thicken the leaflets, as the original thickness was often incompatible with the hardware limitations of 3D printers.

The conversion of the aorta and LV inner wall involved a similar process of mapping a closed surface mesh into segmentation, converting it to an SDF representation, obtaining a new thickness-adjusted segmentation, and performing the component-agnostic meshing operations. The SDF voxelgrid resolution was set at an isotropic 0.5mm spacing.

In addition to these steps, the aorta and LV were combined at two different cardiac phases to design the optimal geometry for our benchtop experiments. More specifically, the systolic aorta and the diastolic LV were combined to accurately model the valve opening and the most relaxed LV structure. From the DeepCarve outputs at both phases, the inner walls of the desired structures at each phase were extracted utilizing the template element information. With a na\"ive combination of those meshes using the canonical coordinate system, the initial merge is almost guaranteed to be inaccurate due to the global movements of those structures throughout the cardiac cycle (Fig.~3C, Initial merge).

Thus, a mesh displacement optimization was further performed to correct for these inconsistencies. The overall idea is to designate a small amount of transition elements in between the aorta and LV and allow those elements to be non-rigidly deformed, while only rigidly transforming the rest of the aorta and LV elements. The following equation was used to maintain a reasonable quality of the transition elements
\begin{equation}
    \mathcal{L} = \lambda_1 \mathcal{L}_{normal} + \lambda_2 \mathcal{L}_{lap} + \lambda_3 \mathcal{L}_{edge}
\end{equation}

\noindent where all loss components were defined previously, and $\lambda_i = \{10,1,10\}$. The deformation field for the transition elements was identical to the deformation fields for the conventional registration and DeepCarve. Separate Adam optimizers were used with learning rates of 1e-3 and 3e-2 for non-rigid and rigid transformations, respectively.

Note that the inner wall mesh was the input to the subsequent segmentation and SDF conversions, so its quality was not critical to the final 3D printing mesh. However, a reasonable quality of the combined mesh was still required to ensure robust performance of the image stencil operation for the intermediate voxelgrid segmentation.

\subsection*{Finite element model of patient-specific anatomies}

Three-dimensional LV, aorta, valve, and calcium geometries were obtained as direct output from the multi-frame meshing approach. The three-dimensional LV and aorta geometry were further extended using solid modeling computer-aided design (CAD) software (SolidWorks 2023, Dassault Systèmes) to improve the flow convergence and stability. The extended inlet connected to LV represented the flow entrance from LA and LV during the cardiac cycle. The generated geometries were imported into Abaqus 2022 software (Simulia, Dassault Systèmes). Nonlinear explicit dynamic analysis was performed to simulate the mechanical response of the FSI analysis. For this study, the modified anisotropic hyperelastic Holzapfel–Gasser–Ogden material model \cite{holzapfel2000new, gasser2006hyperelastic} was adopted to characterize the mechanical behavior of native AV leaflets. Local coordinate systems were defined for each leaflet to include local fiber orientation. We assumed that the mean fiber directions were symmetric with respect to the circumferential axis of the local coordinate system (Fig. S3). Specific material constants were utilised for calcium, as previously outlined in the literature. Isotropic linear elastic model was implemented to characterize the mechanical properties of the calcium deposits. The nodal points of the calcium elements were tied to leaflets at the intersection of the leaflet surface. In the instance of conventional registration, only those leaflet surface areas in contact with calcium were identified and assigned the mechanical properties of the calcium deposits, without being reconstructed three-dimensionally.  
\subsection*{Fluid-Structure Interaction Modeling}

FSI modeling technique was utilized to simulate intravascular hemodynamic interactions with the deformable valve structure. The fluid domain was spatially discreted, and governing equations for both the fluid and structure were solved separately for each discrete time step. The flow patterns in the LV and aorta were simulated using a commercial fluid solver package (XFlow 2022x, Dassault Systèmes), utilizing a Large-Eddy Simulation (LES) turbulence model. The fluid-structure system is represented using the immersed finite element/finite difference method to avoid issues related to mesh motion and mesh regeneration resulting from significant deformations. A method including iterative implicit 1-way and 2-way coupling was used to compute the numerical FSI problems. The 1-way coupling approach is employed to model the cardiac motion for individual patients, where the nodal displacement is prescribed to the LV mesh using our multi-frame image registration framework. An iterative 2-way coupling method is then used to simulate flow interaction between the fluid domain and the valve structure.   The blood was mathematically treated as an incompressible Newtonian fluid with a density of 1050 kg/m3 and a dynamic viscosity of 0.0035 Pa s. The inlet flow boundary condition for this analysis was based on a representative flow waveform derived from healthy adult physiology described in the literature \cite{PanesarMV,milnor1989hemodynamics}. The outlet pressure boundary condition at the aorta was set based on blood pressure readings taken during a pre-TAVR transthoracic echo examination. The duration of the cardiac cycle for each patient was determined by the heart rate measured during the pre-TAVR echo testing. Overall three cardiac cycles were simulated and their results were analyzed, taking into account that the difference in observed physical quantities was less than five between the cycles. The stable time increment was established at 1×10\textsuperscript{-5}s to ensure numerical stability and temporal accuracy. Grid independence was assessed on the first patient model by employing three distinct grid sizes (0.8mm, 0.6mm, and 0.4mm lattice resolution). The stability measure converged with a 4.42\% difference between the medium and fine grids, indicating that the medium grid (0.6 mm resolution, 510,000 elements) was considered to be grid-independent. Each FSI simulation was completed in 30.6 hours (3 cardiac cycles) on a desktop PC with a 3.0 GHZ i7-9700 processor with 8 cores and 32 GB RAM.

\subsection*{Experimental study design}
In vitro validation was conducted on a subset of patients (patients 1 and 2) selected based on differences in their severity of AS. The experimental setup used in this work was analogous to that previously published by our group \cite{rosalia2023soft}.   We employed a custom soft robotic LV sleeve that mimics the beating motion of the heart using 3D printed geometry. The soft robotic LV contracts pneumatically to pump fluid, causing the valve to move during systole. We created the patient's artificial anatomy, specifically the LV sleeve, and conducted hemodynamic evaluation using a hydrodynamic flow loop.  The 3D-printed anatomical model of each patient was combined with the LV sleeve in a hydrodynamic flow loop. Hemodynamic parameters related to AS were assessed using continuous wave and color flow mapping Doppler. Results were compared with the clinical transthoracic echocardiogram data of the patients. 

\subsection*{ Manufacturing of patient-specific heart and sleeve}
The 3D anatomy of the LV, aorta (ascending segment), AV leaflets, and calcium were exported in shell stereolithography (STL) format for each patient using our 3D printing mesh conversion technique. The generated STLs of aortic leaflets and the calcium geometry were integrated with the aorta and LV using a Boolean merging operation. The combined STL file was imported into Preform software (v3.21, Formlabs), and the support material's structure was manually altered to optimize the printability and the amount of internal support material. The LV, aorta, AV leaflets, and calcium were printed simultaneously on a Form 3B STL 3D printer (Formlabs) with a layer thickness of 0.1 mm and a wall thickness of 1.3 mm

A soft robotic LV sleeve made of inflatable Thermoplastic Polyurethane (TPU) pockets wrapped around the LV anatomy was used to actuate the system. Patient-specific sleeves were initially created utilizing computer-aided software (SolidWorks 2023, Dassault Systèmes). The outside surface of the LV was offset by 10 mm to establish a patient-specific shape. The generated shape was then projected onto a flat surface to form the shapes of the molds for production. TPU sheets were formed using the molds and then heat-sealed to make sealed and inflatable pockets integrated into the LV sleeve based on previously described methods by our group \cite{rosalia2023soft,rosalia2024modulating}.

\subsection*{Mock circulatory flow loop}
A mock circulatory loop was created utilizing hydraulic and mechanical components to mimic the blood flow (viscosity of medium, $\mu$\ = 1.0 cP) in the anatomy. The loop for each patient was created by connecting the specific 3D-printed anatomy to soft PVC plastic tubing (5/8-inch inner diameter, 1-inch outer diameter; McMasterr-Carr), two variable-resistance ball valves to simulate arterial and venous resistance, and custom-made acrylic compliance chambers to represent peripheral compliance.
The LV sleeve was connected to a control box with pressure regulators and actuated using input-pressure square waveforms ranging from 0 psi (diastole) to 10 psi (systole), with a duty cycle of 30-50\% and a heart rate of 40-60 bpm. The sleeve was connected in series to adjustable resistance valves and compliance chambers, both on the outflow and the inflow, that could be adjusted to re-create patient-specific hemodynamic states. A unidirectional mechanical valve (Regent 19AGN-751, Abbott Laboratories) was connected to simulate the function of the mitral valve.

A clamp-on flow probe (PS series, Transonic) in the position of the mitral valve and two straight-tip 5F PV catheters advanced to the LV and aortic position were used to monitor pressures and flows. The flow probes were connected to a two-channel flowmeter console (400-series, Transonic), which was connected to an 8-channel Powerlab system (ADInstruments) for data acquisition and recording. The catheters were connected to a Transonic ADV500 PV System and to the Powerlab (ADInstruments). All PV data were processed and analyzed on Matlab R2020a (MathWorks).

\subsection*{Echocardiography}
The Epiq CVx cardiovascular ultrasound system and the X5-1 transducer (Philips) were used for echocardiographic evaluation on the physical prototypes. The transducer was positioned directly on the printed aorta and the ultrasonic beam was aligned with the direction of flow for continuous wave Doppler imaging. CW Doppler was used for the evaluation of v\textsubscript{LVOT}, v\textsubscript{peak} and estimates of dP\textsubscript{max} and dP\textsubscript{mean}. Dimensionless velocity index (DVI) was calculated as the ratio of the LVOT velocity and the maximum velocity obtained by CW Doppler across the aortic valve.
\begin{equation}
DVI=\frac{V\textsubscript{LVOT}}{V\textsubscript{peak}}
\end{equation}
Images of the LV and measurements of EDD, ESD, and d\textsubscript{LVOT}  were obtained in B-mode. Estimates of LV volumes, both at ESV and EDV, were obtained using the Teicholz formula \cite{teichholz1976problems}:
\begin{equation}
LVV=\frac{7}{2.4+LVID}\times{LVID^3}
\end{equation}
where LVV and LVID are the LV volume and internal diameter, respectively.

\bibliography{scibib}

\bibliographystyle{Science}

\section*{Acknowledgments}
We acknowledge funding from multiple organizations, the National Science Foundation (NSF) grant 1847541 (to ETR), the Additional Ventures Single Ventricle Research Fund (to ETR), the National Institute of Health (NIH), National Heart, Lung, and Blood Institute (NHLBI) grant F31HL16250 (to DHP), grants R01HL121226, R01HL142036, and T32HL098069 (to JSD), grants R01EB033853, R01HL151704 and R01HL159010 (to CTN). Parts of some figures were created using BioRender, Adobe Illustrator, and OriginPro 2021b software.

\section*{Author contributions}

CO, DHP, JSD, and ETR designed the concept, CO and DHP performed computational investigation and analysis, CO, LR, MER and DG performed benchtop experiments, RM provided the dataset, CO, DHP, LR, DG, CTN, JSD, and ETR wrote or revised the manuscript,  CTN, JSD and ETR provided funding and supervision.

\section*{Competing interests}

ETR is a member of the board of directors at Affluent Medical and also serves on the board of advisors for Pumpinheart and Helios Cardio. ETR offers consulting services for Holistick Medical and is a co-founder of Spheric Bio and Fada Medical. A provisional patent No. 63/611,903 was filed for the C-MAC algorithm (DHP). The other authors declare that they have no competing interests.

%Here you should list the contents of your Supplementary Materials -- below is an example. 
%You should include a list of Supplementary figures, Tables, and any references that appear only in the SM. 
%Note that the reference numbering continues from the main text to the SM.
% In the example below, Refs. 4-10 were cited only in the SM.     

% Figs. S1 to S3\\

% For your review copy (i.e., the file you initially send in for
% evaluation), you can use the {figure} environment and the
% \includegraphics command to stream your figures into the text, placing
% all figures at the end.  For the final, revised manuscript for
% acceptance and production, however, PostScript or other graphics
% should not be streamed into your compliled file.  Instead, set
% captions as simple paragraphs (with a \noindent tag), setting them
% off from the rest of the text with a \clearpage as shown  below, and
% submit figures as separate files according to the Art Department's
% instructions.

\section*{List of Supplementary Materials}
Fig. S1 to S5

\noindent Movie S1 to S5
\end{document}